\documentclass[
reprint,
superscriptaddress,
showkeys,
amsmath,amssymb,
aps,
prx,
floatfix,
]{revtex4-2}

\usepackage[
activate={true,compatibility},
final,
tracking=true,
kerning=true,
nopatch=footnote]{microtype}

\usepackage{mathtools}
\usepackage{amsfonts}
\usepackage{nicefrac}
\usepackage{physics}
\usepackage{color}
\usepackage{graphicx}
\usepackage{dcolumn}
\usepackage{booktabs}
\usepackage{tabularx}
\usepackage{multirow}
\usepackage{xcolor}
\usepackage[normalem]{ulem} 
\usepackage{amsthm}
\usepackage{tikz}

\usetikzlibrary{positioning}        % <— for “below right=… of …”
\usepackage[margin=1in]{geometry}
\usepackage{subcaption} % modern replacement for subfigure
\usepackage[ruled,vlined]{algorithm2e}
\SetKwInput{KwIn}{Input}
\SetKwInput{KwOut}{Output}

\usepackage[colorlinks=true,citecolor=blue,urlcolor=blue,linkcolor=blue]{hyperref}

\makeatletter
\renewcommand\@makecaption[2]{
  \par
  \vskip\abovecaptionskip
  \begingroup
   \small\rmfamily
    \begingroup
     \samepage
     \flushing
     \let\footnote\@footnotemark@gobble
     \@make@capt@title{#1}{#2}\par
    \endgroup
  \endgroup
  \vskip\belowcaptionskip
}
\makeatother

\newcommand{\pwisein}{\left\{ \begin{array}{ll}}
\newcommand{\pwiseout}{\end{array}\right.}

\begin{document}
%\linenumbers
%\setmathlines

\title{Tensor-network approach to quantum optical state evolution beyond the Fock basis}

\author{Nikolay Kapridov}
\email{kapridov.na@phystech.edu} 
\affiliation{Moscow Institute of Physics and Technology, Dolgoprudny, Moscow Region 141700, Russia}
\affiliation{Russian Quantum Center, Skolkovo, Moscow 121205, Russia}

\author{Egor Tiunov}
\email{egor.tiunov@tii.ae}
\affiliation{Quantum Research Center, Technology Innovation Institute, Abu Dhabi, UAE}

\author{Dmitry Chermoshentsev}
\email{dac@rqc.ru}
\affiliation{Moscow Institute of Physics and Technology, Dolgoprudny, Moscow Region 141700, Russia}
\affiliation{Russian Quantum Center, Skolkovo, Moscow 121205, Russia}

%\date{March 11, 2025}

\begin{abstract}  
Understanding the quantum evolution of light in nonlinear media is central to the development of next-generation quantum technologies. Yet, modeling these processes remains computationally demanding, as the required resources grow rapidly with photon number and phase-space resolution. Here, we introduce a tensor-network approach that efficiently captures the dynamics of nonlinear optical systems in a continuous-variable representation. Using the matrix product state (MPS) formalism, both quantum states and operators are encoded in a highly compressed form, enabling direct numerical integration of the Schrödinger equation. We demonstrate the method by simulating degenerate spontaneous parametric down-conversion (SPDC) and show that it accurately reproduces established theoretical benchmarks—energy conservation, pump depletion, and quadrature squeezing—even in regimes where conventional Fock-basis simulations become infeasible. For high-intensity pump fields  $(\alpha=100)$, the MPS representation achieves compression ratios below $3\times10^{-4}$ while preserving physical fidelity. This framework opens a scalable route to modeling multimode quantum light and nonlinear optical phenomena beyond the reach of traditional methods.
\end{abstract}

\maketitle 

\section{Introduction}
Optical and photonic devices with nonlinear properties are rapidly advancing, finding widespread applications in quantum technologies~\cite{giordani2023integrated, wang2020integrated}. To ensure their effective implementation, it is essential to predict the behavior of optical quantum states under different conditions. However, the simulation of the quantum state dynamics is a challenging task because of the high dimensionality of the Hilbert space~\cite{feynman1982simulating}. This challenge motivates the development of new simulation techniques to limit the rapid growth of required computational resources with the Hilbert-space dimension. Simulating a quantum state’s evolution involves solving the Schrödinger equation, which requires selecting a representation of the state. A commonly accepted strategy is to represent optical quantum states in discrete Fock basis. This basis is infinite, however, it is typically truncated to a finite subspace in practice~\cite{chinni2024beyond, nikitin1991quantum}. Alternatively, wavefunctions can be represented in a continuous basis~\cite{lvovsky2009continuous,fedotova2023continuous}. In the numerical simulation, continuous variables are discretized to the extent necessary to resolve all the details of the encoded wavefunction.

\begin{figure*}[t]
    \centering
    \includegraphics[width=\textwidth]{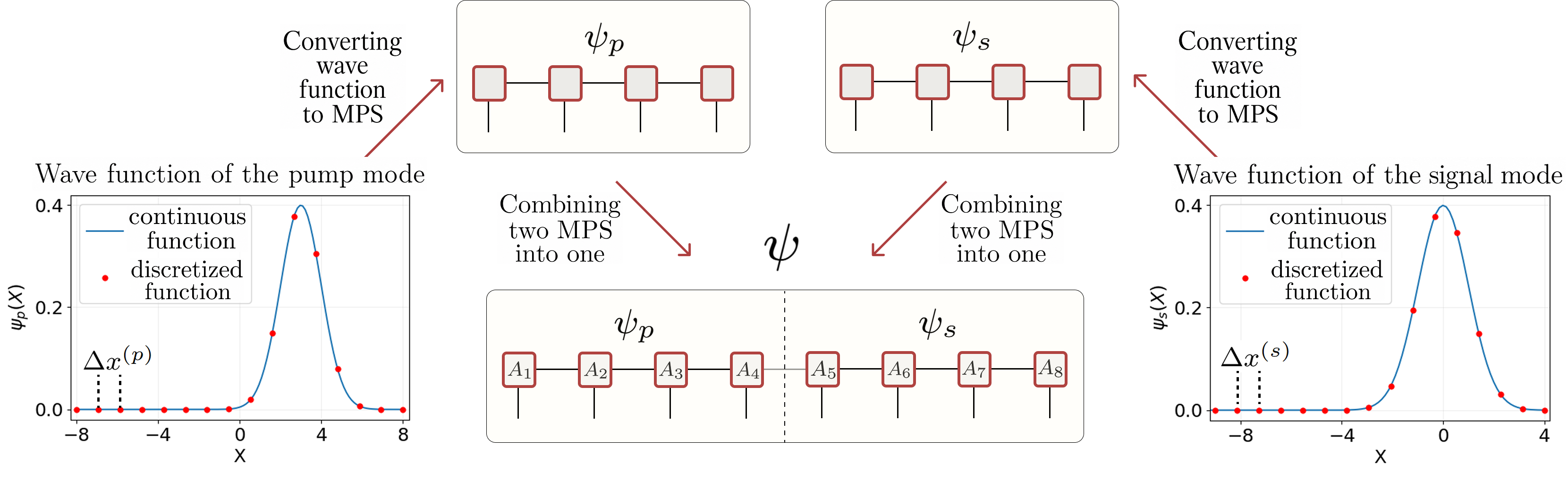} % Замените 'example-image' на путь к вашей картинке
    \caption{Encoding the wavefunction of the entire system in the MPS representation. Wavefunction of each mode is discretized with the required precision. Then it is converted to the MPS format through successive singular value decompositions (SVDs). The MPS representation of the full system is then constructed by concatenating the MPS tensors associated with the individual modes, which corresponds to taking the tensor product of the mode states in the composite Hilbert space. In this example, the wavefunction of the pump mode is defined on the interval \(R^{(p)}=[-8,8]\), while that of the signal mode is defined on the interval \(R^{(s)}=[-9,4]\). Each mode is discretized with N = 16 points which corresponds to a MPS with 4 tensors (for each mode). The MPS of the entire system contains 8 tensors.}
    \label{encoding_wf_to_mps}
\end{figure*}

It turns out that describing continuous wavefunctions with exponentially high resolution is possible with the help of tensor networks. One of the most effective and widely used types of tensor networks is the Matrix Product State (MPS)~\cite{white1992density, white1993density, ostlund1995thermodynamic, rommer1997class, perez2006matrix, dukelsky1998equivalence, verstraete2006matrix}. 
It has proven to be an efficient ansatz for ground states of gapped 1D Hamiltonians~\cite{jaschke2018one, hastings2006solving, pirvu2012matrix}. The MPS representation was independently discovered by the applied math community under the name of Tensor Train (TT)~\cite{oseledets2011tensor, khoromskij2010quantics, lindsey2023multiscale}. Various linear algebraic methods were reimplemented within the TT framework~\cite{oseledets2011tensor, oseledets2012solution}. Several studies have investigated the key properties of grid-discretized functions that allow their efficient representation within tensor-network formats~\cite{khoromskij2011d}. Moreover, a direct connection has been established between the internal complexity of a tensor-network representation and the smoothness of the encoded function~\cite{lindsey2023multiscale,bohun2026entanglement,guzman2026efficient}. Thus, the tensor network approach opens an attractive avenue to simulate high-dimensional continuous problems efficiently and at scale~\cite{gourianov2022quantum,peddinti2024quantum,pisoni2025compression,peddinti2025quantum,kiffner2023tensor,kornev2023numerical,arenstein2025fast}. 
%\textbf{Tensor-network methods have also started to find applications in quantum optics and quantum light propagation problems~\cite{manzoni2017simulating,yanagimoto2021efficient}. However, these approaches were formulated in representations fundamentally different from the continuous quadrature-space framework considered in the present work.}

More recently, MPS-based approaches have also been applied to the simulation of quantum optical processes. In Ref.~\cite{yanagimoto2021efficient}, MPS techniques were used to simulate the propagation of spatially distributed optical pulses containing a small number of photons (\(<4\)) in a strongly nonlinear medium. In Refs.~\cite{manzoni2017simulating,mahmoodian2020dynamics}, the propagation of quantum light through atomic ensembles was studied using an effective spin-model description, where the MPS encoded the states of \(d\)-level atoms or spins.

In this work, we focus on a different problem compared to previously discussed papers. In particular, we simulate quantum dynamics of a nonlinear optical process in high-intensity regime by solving time-dependent Schrödinger equation. When the system is initialized with a large number of photons, for example, when a nonlinear medium is driven by a laser field of high intensity, the corresponding cutoff in the Fock space must be increased to accurately capture all relevant photon-number components. This, in turn, leads to a rapid growth of the Hilbert-space dimension, making direct simulations computationally intensive due to memory and processing-time limitations.

Here, we address this challenge for the degenerate spontaneous parametric down-conversion (SPDC) process~\cite{chinni2024beyond, birrittella2015coherently, scharf1984effect, milburn1981production, couteau2018spontaneous, bandilla2000parametric} by combining the continuous-variable formulation of its quantum evolution with the MPS representation. Accurate resolution of the wavefunction’s fine structure—rapid phase oscillations, interference fringes, and non-Gaussian tails—requires a large phase-space window and a finely sampled grid, resulting in a large discretized state space. To keep the computations tractable, we employ the MPS representation, which compactly factorizes the state while preserving the essential correlations. We validate the accuracy of our approach by simulating the SPDC process with the pump mode initialized in a coherent state $\alpha = 10$ in the Fock basis using conventional sparse linear-algebra methods. We then extend the analysis to a higher amplitude $\alpha = 100$, which more closely reflects experimental conditions where a laser mode typically contains thousands to millions of photons within its coherence volume. To validate the proposed algorithm, we introduce a set of benchmark metrics—total energy conservation, pump depletion, variance of the $X$-quadrature, and numerical error analysis. Since direct benchmarking with the full Fock basis is computationally infeasible for $\alpha=100$, these metrics serve as reliable indicators of simulation accuracy. According to these criteria, the MPS-based simulations faithfully reproduce the expected quantum dynamics, confirming the validity of the proposed approach.

%\textbf{Connection with previous works:}
%Here we briefly summarize previous tensor-network approaches to the simulation
%of quantum-optical processes and clarify how our method differs from them.
%In Ref.~\cite{yanagimoto2021efficient}, MPS techniques were used to simulate
%the propagation of spatially distributed optical pulses containing a small
%number of photons (\(<4\)) in a strongly nonlinear medium. In
%Refs.~\cite{manzoni2017simulating,mahmoodian2020dynamics}, the propagation of
%quantum light through atomic ensembles was studied using an effective spin-model
%description, where the MPS encoded the states of \(d\)-level atoms or spins.

%Our work addresses a different physical setting and state
%representation. We simulate the nonlinear interaction of a small number of
%optical modes in the high-intensity regime using a continuous-variable
%quadrature representation. In this approach, the MPS directly encodes the
%quadrature-space wavefunction of all optical modes rather than a chain of
%spatial photonic bins or atomic spin degrees of freedom.

\section{Methods}
In this paper, we tackle the problem of quantum state evolution governed by the Hamiltonian of the SPDC process in the continuous quadrature basis. In the interaction picture, the Hamiltonian is expressed in terms of photon creation ($\hat{a}^\dagger$) and annihilation ($\hat{a}$) operators as follows:
\begin{equation}
\hat{H} =  i \kappa(\hat{a}_{p}^{\dagger}\hat{a}_{s}\hat{a}_{s} - \hat{a}_{p}\hat{a}_{s}^{\dagger}\hat{a}_{s}^{\dagger}).
\label{eq:hamiltonian}
\end{equation}

Here, $\kappa$ is the coupling constant, which is proportional to the second-order nonlinear susceptibility $\chi^{(2)}$. The subscripts $p$ and $s$ denote the pump and signal modes, respectively. In the quadrature representation, the creation and annihilation operators are given by
\begin{equation}
\left\{
\begin{aligned}
    &\hat{a}^{\dagger} = \dfrac{X-d/dX}{\sqrt{2}},\\
    &\hat{a} = \dfrac{X+d/dX}{\sqrt{2}}.
\end{aligned}
\right.
\label{eq:system}
\end{equation}

To simulate the quantum evolution of the system, we numerically solve the time-dependent Schrödinger equation using the implicit Euler method:
\begin{equation}
(\hat{I}+i\Delta t \hat{H})\psi_{t+1}(x) - \psi_{t}(x) = 0,
\label{eq:diffequation}
\end{equation}
where $\hat{I}$ stands for the identity matrix, $\Delta t$ is the time step, $\psi_t(x)$ is the wavefunction in the position $x$ at time $t$.

The Schrödinger equation \eqref{eq:diffequation} defines a linear system for the wavefunction at time step $t+1$. The signal ($s$) and pump ($p$) components of the wavefunction are discretized on predefined grids of size $N^{(s/p)}$ and resolution $\Delta x^{(s/p)}$ (see Fig.\ref{encoding_wf_to_mps}). The grid spacing is chosen to resolve all fine-scale features of the continuous wavefunction throughout the entire evolution. The spatial support of each component, $[x_a^{(s/p)}, x_b^{(s/p)}]$, is selected such that the wavefunction remains negligible outside this interval at all times. To facilitate the encoding of wavefunctions and operators in MPS format, the grid size is required to be a power of two, i.e., $N = 2^m$.

%must satisfy
%\[
%N_{(s/p)} = 1 + %\frac{x_b^{(s/p)} - x_a^{(s/p)}}%{\Delta x_{(s/p)}},
%\]
%and

We consider the pump and the signal modes as a closed quantum system. Consequently, the corresponding wavefunction captures all information and correlations between the two modes in a single vector of size $N^{(p)} \times N^{(s)}$. The initial state of the system, denoted by $\psi_0$, is taken as a tensor product of a vacuum state for the signal mode, $\psi_0^{(s)}(x)$, and a coherent state for the pump mode, $\psi_0^{(p)}(x)$, respectively:
\begin{equation}
\psi_0^{(s)}(x) = \pi^{-1/4}\exp{(-x^2/2)},
\label{eq:swf}
\end{equation}
\begin{equation}
\psi_0^{(p)}(x) = \pi^{-1/4}\exp{(-(x-x_0)^2/2)}.
\label{eq:pwf}
\end{equation}

Here $x_0 = \alpha \sqrt{2}$, where $\alpha$ is the amplitude of the coherent state.

The novelty of our approach is encoding both the Hamiltonian and the overall wavefunction of the system in the MPS representation~\cite{schollwock2011density, oseledets2011tensor} and solving the Schrödinger equation written in the quadrature basis directly within this framework. To enable this, we begin by expressing the discretized support of the wavefunction in binary representation. Specifically, each grid point $x_k$ is uniquely indexed by a binary string $(i_1, i_2, \dots, i_m)$, where $i_j \in {0, 1}$. This allows us to rewrite the wavefunction as
\begin{equation}
    \ket{\psi} = \sum_k \psi(x_k) \ket{x_k} \Delta x = \sum_{i_1...i_m} \psi_{i_1...i_m} \ket{i_1...i_m}.
\label{eq:binary}
\end{equation}

Afterwards, we utilize the MPS as an ansatz to approximate the tensor of complex amplitudes $\psi_{i_1\cdots i_m}$ in a compressed way:
\begin{equation}
\psi_{i_1\cdots i_m} = \sum_{\alpha_0, \ldots, \alpha_m}A_{\alpha_0i_1\alpha_1}^{[1]} A^{[2]}_{\alpha_1i_2\alpha_2} \ldots A^{[m]}_{\alpha_{m-1}i_m\alpha_m}
\label{eq:mps}.
\end{equation}
Here $A^{[j]}_{\alpha_{j-1}i_j\alpha_j}$ are elements of a three-dimensional tensor $A^{[j]}$ of shape $(\chi_{j-1}, 2, \chi_j), \ \chi_0=\chi_m=1$. Index $i_{j}$ of tensor $A^{[j]}$ matches the corresponding bit in the binary notation (\ref{eq:binary}) and is responsible for selecting one of the two matrices $\lbrace A^{[j]}_{0}, A^{[j]}_{1}\rbrace$ of shape $(\chi_{j-1}, \chi_j)$. In notation (\ref{eq:binary}), each bit value $i_j$ dictates the behavior of the wavefunction across distinct scales: the index $i_1$ selects between the first or the second half of the wavefunction support, while the index $i_2$ splits the first selected half into two halves again, and so on down to finer resolutions. In the MPS representation, this multiscale structure is encoded in the corresponding tensors. Fixing the value of index $i_j$ for each tensor $A^{[j]}$ and multiplying the corresponding matrices allows to compute one element of the original wavefunction. The sizes $\chi_j$ of the matrices $A^{[j]}_{\alpha_{j-1}i_{j}\alpha_j}$, called bond dimensions, determine how distinct scales are correlated between each other.

To encode the initial wavefunction of the full system as a tensor network, we convert the dense discretized wavefunctions of the pump, Eq.~\eqref{eq:pwf}, and signal, Eq.~\eqref{eq:swf}, into MPS form through successive singular value decompositions (SVDs)~\cite{schollwock2011density}. These two MPS representations are then concatenated to form the composite initial state of the full system at $t=0$, given by the tensor product of the pump and signal states in the enlarged tensor-product Hilbert space (see Fig.\ref{encoding_wf_to_mps}). The system Hamiltonian \eqref{eq:hamiltonian} is expressed as a sum of tensor products of creation and annihilation operators, $\hat{a}$ and $\hat{a}^{\dagger}$. Since each of these operators can be efficiently represented as a matrix product operator (MPO), the full Hamiltonian is constructed by combining the MPOs corresponding to different modes via appropriate tensor products (see Appendix A). Furthermore, elementwise products and linear combinations of MPOs can also be carried out efficiently within the MPS formalism~\cite{schollwock2011density}. Once both the operator $\hat{U} \equiv (\hat{I} + i \Delta t \hat{H})$ and the wavefunction $\psi(x)$ are expressed in MPS form, the linear system
\begin{equation}
\hat{U} \psi_{t+1} = \psi_t
\end{equation}
arising from the time-stepping scheme \eqref{eq:diffequation} can be solved using local projection method~\cite{dolgov2014alternating,oseledets2012solution,rohrig2025performance} (see Appendix B for the details).

The MPS representation of continuous, smooth functions is not standard practice in the quantum-optics community, but several heuristic arguments suggest that such a representation naturally admits a low-rank structure. First, 1D smooth functions exhibit universal behavior at fine scales: past a certain bond, the entanglement decays exponentially~\cite{bohun2026entanglement,lindsey2023multiscale}. In the multidimensional case, there is much more freedom in how such functions can be encoded as an MPS. Here we discretize a 2D function in binary form, placing the bits of the first and second arguments next to each other (sequential encoding):
\begin{equation}
    \psi(x_p,x_s) = \sum_{i=1}^N \psi_i(x_p) \phi_i(x_s).
\end{equation}
The entanglement between the first and second sets of bits reflects how separable the function is, quantified by $N$ in the formula above. The bond dimension within each set, in turn, reflects the smoothness of the functions $\psi_i(x_p)$ and $\phi_i(x_s)$. Thus, the expectation of low-rank structure in our setting rests on the joint assumptions of separability and smoothness. In general, low-rank structure of smooth multidimensional functions has been analyzed in the recent tensor train interpolation paper~\cite{guzman2026efficient}.

\section{Results}

\begin{figure*}[t]
    \centering
    \includegraphics[width=1\linewidth]{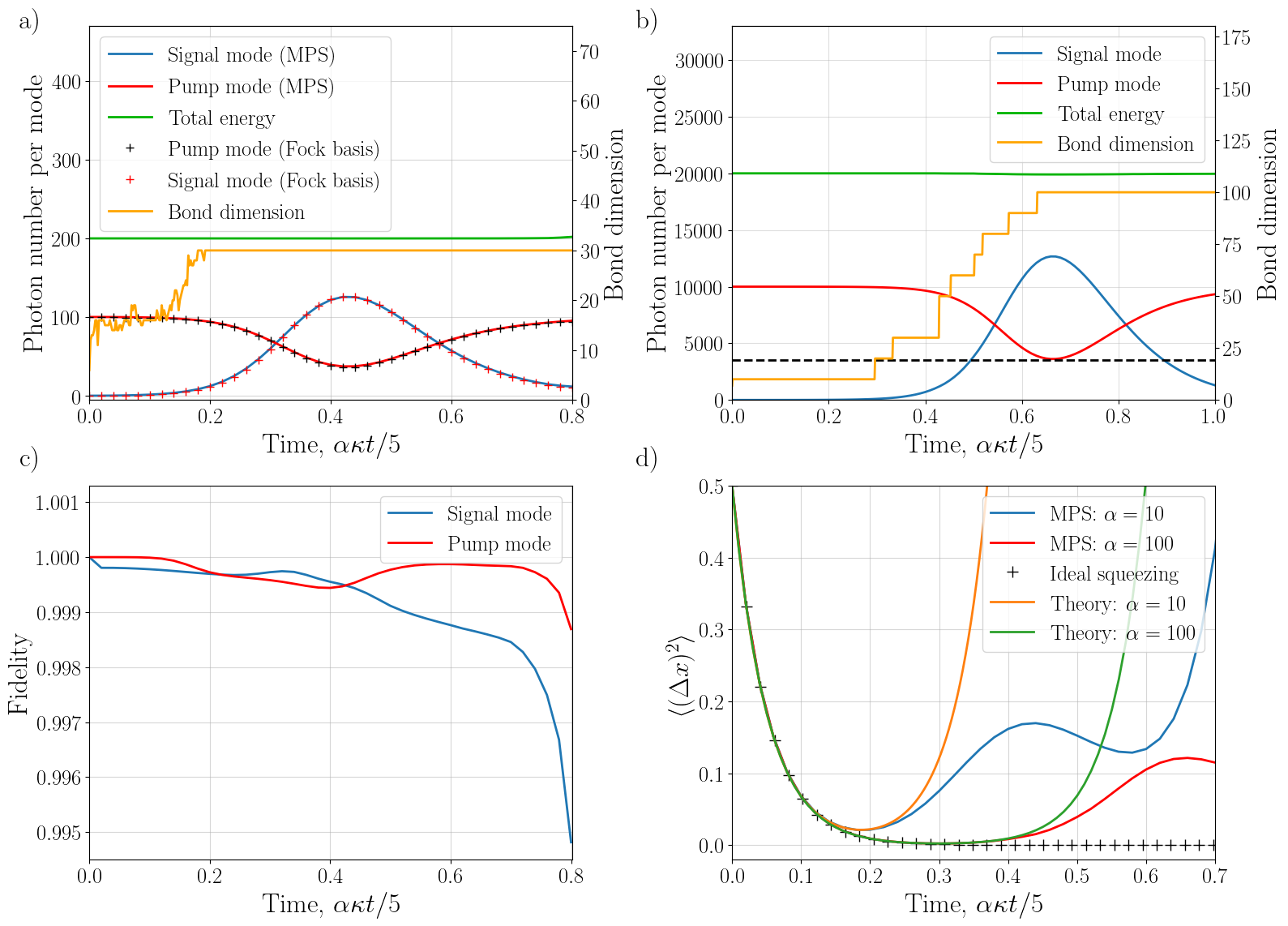}
    \caption{\textbf{(a,b)} Photon population dynamics $\langle n\rangle=\text{Tr}\left(\rho a^\dagger a\right)$ in the pump and signal modes (left Y-axis) and maximum bond dimension ($\chi_{max}=max_{i}(\chi_i)$) change (right Y-axis) during quantum evolution in the SPDC process. Blue and red solid lines correspond to the MPS algorithm; the green line shows the quantity \(2n_p + n_s\), which is proportional to the total energy of the system; orange solid line demonstrates how the maximum bond dimension changes during the evolution. \textbf{(a)} Pump mode is initialized in the coherent state $\alpha = 10$. Crossed dots demonstrate state vector simulation in Fock basis. \textbf{(b)} Pump mode is initialized in the coherent state $\alpha = 100$. Black dashed line shows the theoretical limit of pump mode depletion. \textbf{(c)} Fidelity between Fock-basis state-vector and continuous-basis MPS simulations of the signal and pump density matrices during the quantum evolution for $\alpha=10$. Fidelity calculation is done in the Fock basis. \textbf{(d)} Variance of $X$-quadrature in the signal mode during quantum evolution with the pump mode initialized at $\alpha=10$ and $\alpha=100$. The blue and red lines demonstrate MPS approach, while orange and green lines correspond to the theoretical analysis given in Ref.~\cite{kinsler1993limits}. The black crosses indicate variance for the ideal single-mode squeezing in the parametric approximation (pump mode is treated as a classical field). In panel~\textbf{(d)}, the time axis is rescaled as \(\alpha\kappa t/5\) to align the curves obtained for different values of \(\alpha\) and to make the horizontal axis independent of the particular value of \(\kappa\). The same normalization is used consistently throughout all figures.}
    \label{photon_dynamics_both_cases}
\end{figure*}

\begin{figure*}[ht!]
    \centering
    \includegraphics[width=1\linewidth]{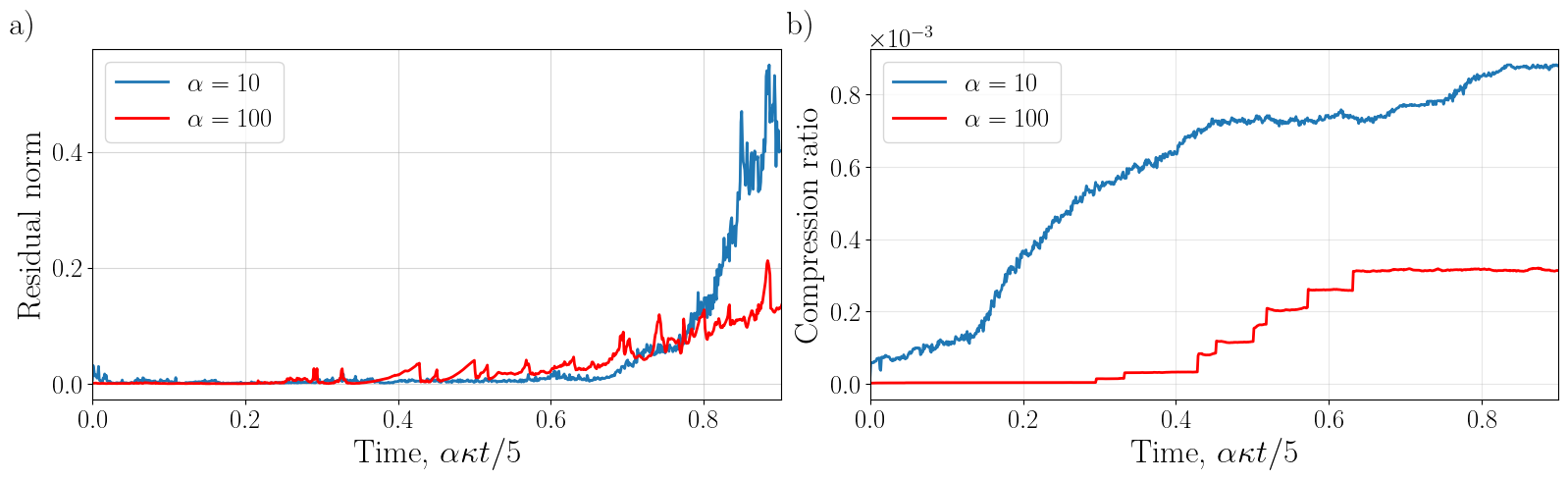}
    \caption
    {\textbf{(a)} $l_2$ residual norm $||\tilde{\psi_t}-U\tilde{\psi}_{t+1}||/||\tilde{\psi}_t||$ of the linear system (\ref{eq:diffequation}) during the quantum evolution simulation. \textbf{(b)} Compression ratio. The ratio between the total number of elements in MPS and the number of elements in dense vector representation ($2^{m_{\mathrm{MPS}}}$) throughout the SPDC evolution. Blue (red) line corresponds to the case when the pump mode is initialized in a coherent state $\alpha = 10$ ($\alpha = 100$).}
    \label{numerical_error_and_compression_ratio}
\end{figure*}

%\begin{figure*}[t]
%    \centering
%    \includegraphics[width=1\linewidth]{Figures/four_plots_on_one_figure__changed_5.png}
%    \caption{\textbf{(a,b)} Photon population dynamics $\langle n\rangle=\text{Tr}\left(\rho a^\dagger a\right)$ in the pump and signal modes (left Y-axis) and bond dimension change (right Y-axis) during quantum evolution in the SPDC process. Blue and red solid lines correspond to the MPS algorithm; green line shows total energy of the system expressed in the number of signal photons; orange solid line demonstrates how the bond dimension changes during the evolution. \textbf{(a)} Pump mode is initialized in the coherent state $\alpha = 10$. Crossed dots demonstrate state vector simulation in Fock basis. \textbf{(b)} Pump mode is initialized in the coherent state $\alpha = 100$. Black dashed line shows the theoretical limit of pump mode depletion. \textbf{(c)} Fidelity between Fock-basis state-vector and continuous-basis MPS simulations of the signal and pump density matrices during the quantum evolution. Fidelity calculation is done in the Fock basis. \textbf{(d)} Variance of $X$-quadrature in the signal mode during quantum evolution with the pump mode initialized at $\alpha=10$ and $\alpha=100$. The blue and red lines demonstrate MPS approach, while orange and green lines correspond to the theoretical analysis given in Ref.~\cite{kinsler1993limits}. The black crosses indicate variance for the ideal single-mode squeezing in the parametric approximation (pump mode is treated as a classical field).}
%    \label{photon_dynamics_both_cases}
%\end{figure*}

To validate our framework, we compare two approaches: (i) Fock-basis simulations using the standard state-vector representation and (ii) continuous-variable simulations employing an MPS ansatz. The pump mode is initialized in a coherent state with amplitude $\alpha = 10$, which requires a Fock-basis cutoff of about 150 photons, which is still tractable for state-vector methods. 
\begin{table}[h!]
\centering
\begin{tabular}{|c|c|c|}
\hline
\textbf{} & \textbf{$\alpha=10$} & \textbf{$\alpha=100$} \\ \hline
$R_s$ & [-10,10] & [-10,10] \\ \hline
 $R_p$ &  [-24,24] &  [-151,151] \\ \hline
 $m_p$ &  12 &  15 \\ \hline
 $m_s$ & 13 &  15 \\ \hline
 $m_{\mathrm{MPS}}=m_p+m_s$ & 25 &  30 \\ \hline
 $x_0=\sqrt{2}\alpha$ & 14.1 &  141 \\ \hline
 $\Delta t$ & 0.001 & 0.0001 \\ \hline
 $\chi_{max}$ & 30 & 100 \\ \hline
\end{tabular}
\caption{Parameters used for the tensor network simulations: $R_s$ ($R_p$) denotes the spatial support of the signal (pump) mode, $m_s$ ($m_p$) denotes the number of bits in the signal (pump) mode, $m_{\mathrm{MPS}}$ denotes the number of bits in the overall wavefunction, $x_0$ is the mean coordinate of the coherent state at the initial time, and $\Delta t$ and $\chi_{\max}$ denote the timestep and the maximum bond dimension used in the simulations, respectively.}
\label{tab:parameters}
\end{table}
For the tensor-network simulation, the signal and the pump modes are initialized with the parameters from the first ($\alpha=10$) column of the Table \ref{tab:parameters}.

Fig.\ref{photon_dynamics_both_cases}~\textbf{(a)} demonstrates photon dynamics in pump ($\langle n_{p}\rangle = \text{Tr}(\rho_p a_p^{\dagger}a_p)$) and signal ($\langle n_{s}\rangle = \text{Tr}(\rho_s a_s^{\dagger}a_s)$) modes during evolution, obtained from the numerical simulations in both the Fock basis and the continuous representation using MPS (see red and blue curves, respectively). %We observe complete coincidence 
The two approaches yield identical results throughout the entire evolution. The correctness of the MPS-based simulations is also confirmed by the conservation of the quantity $2n_p+n_s$ which is proportional to the total energy of the system, which corresponds to the green line in  Fig.\ref{photon_dynamics_both_cases}~\textbf{(a)}. Fig.\ref{photon_dynamics_both_cases}~\textbf{(a)} also demonstrates the maximum bond dimension  ($\chi_{max}=max_{i}(\chi_i)$) of MPS throughout the quantum evolution (orange line). It grows up to 30, which is set as the truncation cutoff. This value is enough since the fidelity between the density matrices obtained using CV-MPS method and the exact Fock-basis method is more than 99.4\% for both signal and pump modes (see Fig.\ref{photon_dynamics_both_cases}~\textbf{(c)}). 

%\begin{figure*}[ht!]
%    \centering
%    \includegraphics[width=1\linewidth]{Figures/combined_figure_with_labels.png}
%    \caption
%    {\textbf{(a)} $l_2$ residual norm $||\tilde{\psi_t}-U\tilde{\psi}_{t+1}||/\sqrt{\mathcal{N}}$ of the linear system (\ref{eq:diffequation}) during the quantum evolution simulation, for pump mode initialized at $\alpha=10$ ($\mathcal{N}=1$) and $\alpha=100$ ($\mathcal{N}=2^5$). \textbf{(b)} Inverse compression ratio. The ratio between the total number of elements in MPS and the mesh size throughout the SPDC evolution. Blue (red) line corresponds to the case when the pump mode is initialized in a coherent state $\alpha = 10$ ($\alpha = 100$).}
%    \label{numerical_error_and_compression_ratio}
%\end{figure*}

Then we switch to the regime where the pump mode is initialized in a coherent state with amplitude $\alpha  = 100$. Computation in the Fock basis is infeasible on our hardware (Intel Core i7-10700 CPU @ 2.90\,GHz, 32\,GB RAM) even when using sparse linear algebra operations, since it requires solving a linear system in a vector space of dimension around $10^8$ for each timestep. 
%\et{make a table for all parameters}
For this scenario, we adjusted all parameters of the signal and the pump modes according to the second ($\alpha=100$) column of the Table~\ref{tab:parameters}.

Fig.\ref{photon_dynamics_both_cases}\textbf{(b)} (red and blue lines) shows photon-number dynamics in the pump and signal modes during quantum evolution. In this case, we have no direct reference to compare our results with, for the reasons discussed in the previous paragraph. Therefore, to validate our algorithm, we present the self-convergence analysis in respect to key numerical parameters (see Appendix C) and four figures of merit: total energy conservation, benchmarks for pump depletion, variance of $X$-quadrature, and numerical error analysis. We begin with the energy conservation: since we are modeling a closed quantum system, we make sure that the total energy of the system is conserved as demonstrated in the Fig.\ref{photon_dynamics_both_cases}\textbf{(b)} by the green line. Next, theoretical studies of the SPDC process report an energy transfer limit from the pump to the signal mode of about 65–66\%~\cite{bandilla2000parametric, fleischhauer1999long}. The black horizontal dotted line in the Fig.\ref{photon_dynamics_both_cases}\textbf{(b)} represents this limit and closely matches the maximum depletion of the pump mode observed in our simulation. In addition, the variance of the squeezed quadrature for different initial pump-mode amplitudes has been theoretically derived using perturbation theory in previous studies~\cite{kinsler1993limits, hillery1984path, crouch1988limitations}. The corresponding theoretical curves for pump amplitudes 
$\alpha=10$ and $\alpha=100$ are shown in the Fig.\ref{photon_dynamics_both_cases}\textbf{(d)} in orange and green, respectively. These curves are determined by the two leading terms of the perturbation series and are valid while $\exp{(2\alpha \kappa t)}\ll16\alpha^2/3$ (see Ref.~\cite{kinsler1993limits}), which include the point of maximum quadrature squeezing. They coincide with our computations at least up to this point. Finally, in  Fig.\ref{numerical_error_and_compression_ratio}\textbf{(a)}, we compute the $l_2$ residual norm of the linear system $||\tilde{\psi_t}-U\tilde{\psi}_{t+1}||/||\tilde{\psi}_t||$, as a function of time. The error observed for $\alpha = 100$ is quantitatively similar to that for $\alpha = 10$. Since our calculations match with the simulation in Fock basis for $\alpha = 10$, this fact supports the reliability of the results obtained for the higher value of $\alpha$.

The computation with the pump mode initialized at $\alpha=100$ becomes feasible due to the compression capability of the MPS ansatz. To quantify it, we compute the compression ratio at each step of the quantum evolution (see Fig.\ref{numerical_error_and_compression_ratio}\textbf{(b)}). This ratio is defined as the total number of elements in the MPS tensors divided by the number of elements in the corresponding dense-vector representation, \(2^{m_{\mathrm{MPS}}}\). 
%It displays the compression efficiency of MPS compared to the standard dense vector representation. 
Throughout the entire evolution, this ratio remains below $(3\times10^{-4})$ for $(\alpha=100)$ and below $(10^{-3})$ for $(\alpha=10)$.

\section{Conclusion and outlook}
We present a novel numerical approach for modeling the evolution of quantum optical states. Our method uses a discretized continuous representation of states and operators, which are then encoded in the MPS format. The entire algorithm operates within the MPS framework, eliminating the need for going back to the dense vector representation. To validate our approach, we compared quantum evolution simulations of the SPDC process in the Fock basis and in the continuous representation using tensor networks. Both approaches match each other in a low-amplitude regime ($\alpha = 10$). Remarkably, our technique enables simulations of quantum states containing thousands of photons per mode, a regime where traditional methods face significant challenges. Such challenging regime was explored for pump amplitude $\alpha=100$.
%In the absence of direct benchmarks for $\alpha=100$,
The accuracy of the MPS results is assessed through a self-convergence analysis and further supported by several sanity checks, including energy conservation, consistency with previous theoretical studies, the observed level of pump depletion, and the numerical residuals associated with the solution of the linear systems.

This work lays the groundwork for subsequent studies aimed at simulating quantum effects in optical ring microresonators~\cite{vernon2015strongly, strekalov2016nonlinear}. In particular, future efforts will focus on modeling multimode quantum optical evolution in media with $\chi^{(2)}$ and $\chi^{(3)}$ nonlinearities~\cite{raja2019electrically, herr2012universal, tatarinova2025optimization, vorobyev2025optimization}. While the current framework provides an effective method for simulating quantum evolution, further refinement of the time-stepping scheme is needed to scale up the technique to multiple modes. One of the promising directions here is time dependent variational principle formulated in MPS format~\cite{haegeman2011time, lubich2015time, haegeman2016unifying}. Additionally, branched tensor-network architectures, such as the Tucker representation~\cite{de2000multilinear,tucker1966some} or tree tensor networks~\cite{shi2006classical}, may be particularly suitable for multimode optical states. Their hierarchical structure could assign a separate MPS branch to each optical mode and use the upper levels of the network to encode intermode correlations. 

These developments would move MPS-based simulation toward a practical, scalable tool for multimode nonlinear quantum optics, a regime that remains largely out of reach for conventional methods.

\section*{Acknowledgment}
The work was supported by the Russian Science Foundation (Grant No. 25-12-00263)

\bibliographystyle{SciPost_bibstyle}
\bibliography{mybibl}

\clearpage

\section{Appendix} 
\subsection{Constructing Hamiltonian MPO}
To derive an MPO of the SPDC Hamiltonian, we first rewrite it in the quadrature representation. The creation and annihilation operators are expressed in terms of quadrature operators and derivatives according to Eq.(\ref{eq:system}). The SPDC Hamiltonian then takes the form

\begin{equation}
\begin{aligned}
\hat{H} = -i\dfrac{\kappa}{\sqrt{2}}(\dfrac{d}{dX_p}\otimes(X_s^2 + \dfrac{d^2}{dX_s^2}) -\\ -X_p\otimes(1_s + 2X_s\dfrac{d}{dX_s})).
\end{aligned}
\end{equation}

To construct the MPO of the Hamiltonian, we first introduce 
MPO of the elementary operators \(X\), \(X^2\), 
\(d/dX\), and \(d^2/dX^2\). The complete Hamiltonian is then assembled 
directly in MPO form using standard linear-algebra operations, including 
operator addition, operator multiplication, and multiplication by scalar 
coefficients~\cite{schollwock2011density}. Tensor products of operators 
acting on different modes are represented by concatenating the corresponding MPOs.

The position operator $\hat{X}$ for the interval $[x_a, x_b]$ is discretized with a step size $\Delta x = (x_b - x_a)/(2^n-1)$, has the following matrix representation:

\begin{equation}
X = \begin{pmatrix}
x_1 & 0 & \cdots & 0 \\
0 & x_2 & \cdots & 0 \\
\vdots & \vdots & \ddots & \vdots \\
0 & 0 & \cdots & x_N
\end{pmatrix},
\end{equation}
where \(x_k = x_a + (k - 1)\Delta x \) for \( k = 1, 2, \dots, N = 2^n \).
It can be analytically shown that this matrix can be represented as an MPO:
\begin{equation}
X = A_1 A_2 \cdots A_{n-1} A_{n},
\label{eq:x_mpo}
\end{equation}
where
\begin{equation}
A_1=
\begin{pmatrix}
    \begin{pmatrix}
        1 & 0 \\
        0 & 1
    \end{pmatrix}
    &
    \begin{pmatrix}
        0 & 0 \\
        0 & 1
    \end{pmatrix}
\end{pmatrix},
\end{equation}
\begin{equation}
A_2,\cdot\cdot\cdot,A_{n-1} = 
\begin{pmatrix}
    \begin{pmatrix}
        1 & 0 \\
        0 & 1
    \end{pmatrix}
    &
    \begin{pmatrix}
        0 & 0 \\
        0 & 1
    \end{pmatrix}
    \\
    \begin{pmatrix}
        0 & 0 \\
        0 & 0
    \end{pmatrix}
    &
    \begin{pmatrix}
        2 & 0 \\
        0 & 2
    \end{pmatrix}
\end{pmatrix},
\end{equation}
\begin{equation}
A_{n} = 
\begin{pmatrix}
    \begin{pmatrix}
        x_a & 0 \\
        0 & x_a + \Delta x
    \end{pmatrix}
    \\
    \begin{pmatrix}
        2\Delta x & 0 \\
        0 & 2\Delta x
    \end{pmatrix}
\end{pmatrix}.
\end{equation}

The inner matrices of the size 2x2 correspond to physical dimension. The outer matrix which contains smaller ones as its elements correspond to bond dimension.

For the derivative operators \(D^{(1)}\) and \(D^{(2)}\), which are approximated using second-order central differences, and for the \(X^2\) operator, similar expressions can be written. The corresponding MPOs have a constant bond dimension of 3.

\subsection{Solving systems of linear equations within
MPS representation}

\begin{figure*}[t]
    \centering
    \includegraphics[width=\textwidth]{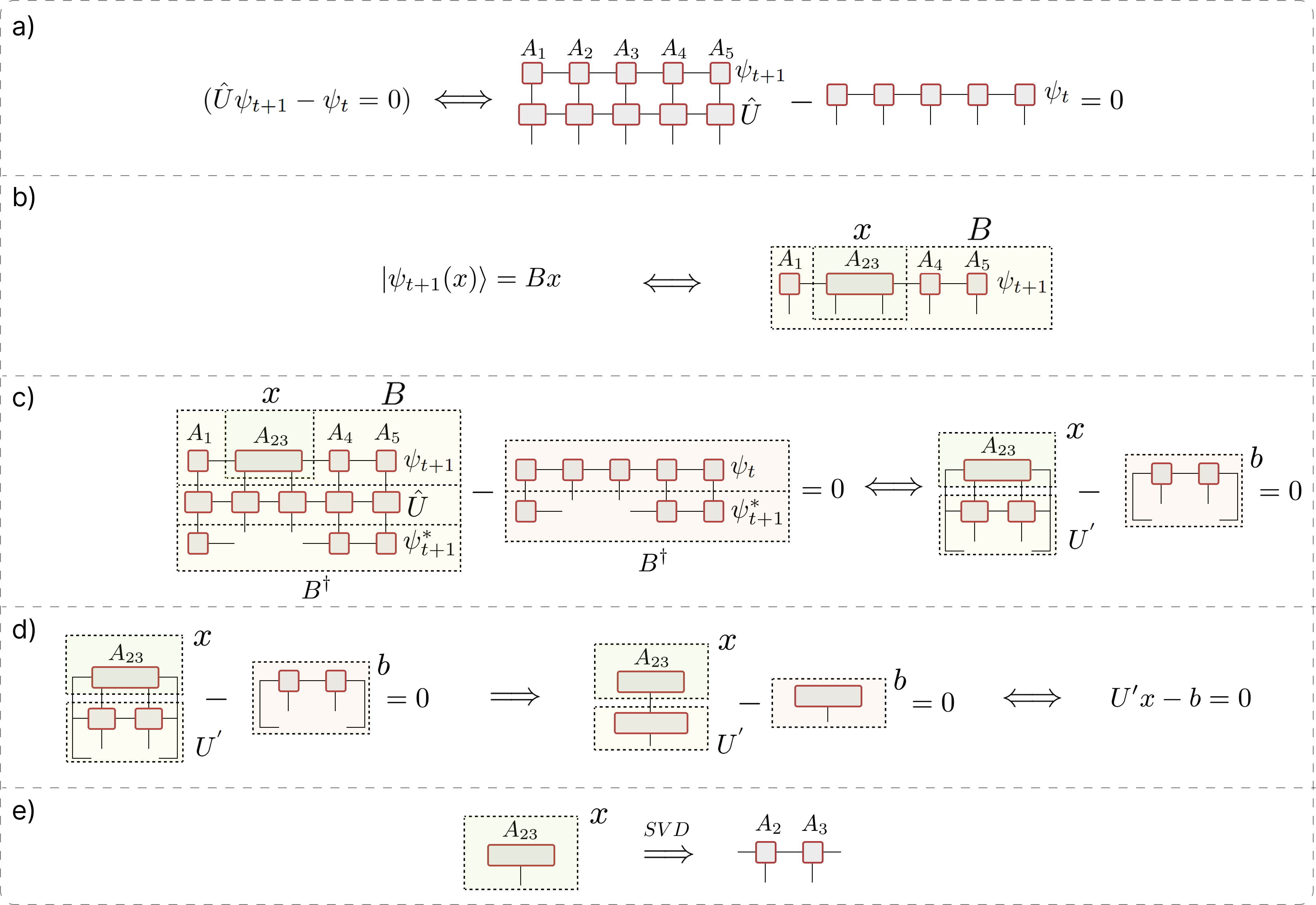}
    \caption{Diagrammatic illustration of the projection algorithm for
solving a linear system in the MPS format. 
(a) Global linear system represented in MPS/MPO form. 
(b) Embedding of the local two-site variable \(x\) into the full trial state
\(\ket{\psi_{t+1}(x)}\). 
(c) Projection of the global equation onto the local variational space. 
(d) Reduced local linear system for the optimized tensor pair. 
(e) Decomposition of the optimized local tensor \(x\) into two MPS tensors.}
    \label{solving_linear_sys}
\end{figure*}

At each time step, we solve the following linear system:
\begin{equation}
U\psi_{t+1} = \psi_{t},
\qquad
U=I+i\Delta t H .
\end{equation}

We solve it iteratively by projecting the operator \(U\) and the right-hand side
$\psi_t$ onto a sequence of lower-dimensional local subspaces and
finding the corresponding local approximations within these subspaces~\cite{saad2003iterative}.
The MPS structure provides a natural way to construct such local subspaces~\cite{oseledets2012solution} (the main computational steps of the MPS solver are shown in Fig.~\ref{solving_linear_sys}). 
Sequential optimization of all tensors is called a sweep. Sweeping over all neighboring tensors yields an ALS/MALS-type iterative solution of the time-stepping equation~\cite{rohrig2025performance,dolgov2014alternating}. Since \(U\) in our task is non-Hermitian, this procedure should not be interpreted as the minimization of a Hermitian energy functional. Instead, it is a local projection linear solver, which does not require \(U\) to be Hermitian~\cite{saad2003iterative,oseledets2012solution}. Such an approach has been successfully applied to linear systems with nonsymmetric matrices~\cite{dolgov2014alternating}.

\subsection{Self-convergence study}

In our numerical simulations, we conduct self-convergence analysis in the case of $\alpha=100$ for the following parameters: time-step $\Delta t$, signal and pump mode quadrature resolution $\Delta x$, and maximum bond dimension $\chi_{max}$. To assess convergence, we selected three representative time moments,
\(t_1 = 0.34\), \(t_2 = 0.46\), and \(t_3 = 0.64\), and performed short
simulations over a time interval \(T_{test} = 0.01\) while varying the numerical parameters and monitoring the full state fidelity error
\(1-F(q,q_{\mathrm{ref}})\), where

\begin{equation}
F(q,q_{\mathrm{ref}})
=
\left|
\langle
\psi_q
|
\psi_{q_{\mathrm{ref}}}
\rangle
\right|^2.
\end{equation}

Here \(q\in\{\Delta t,\Delta x,\chi_{max}\}\) denotes the
varied parameter and \(q_{\mathrm{ref}}\) is the corresponding reference value
with the finest resolution. These time slices represent physically distinct stages of the evolution: the signal mode exhibits maximal quadrature squeezing at $t_1$; approximately \(10\%\) of the pump energy has been
converted into the signal mode at $t_2$; and the energy conversion to
the signal mode is close to its maximum at $t_3$.
For each varied parameter \(q\), we evaluate infidelity $1-F_q(T)$ at $T \in \{t_1+T_{\mathrm{test}},\, t_2+T_{\mathrm{test}},\, t_3+T_{\mathrm{test}}\}$.

In this section, the time values $(t_1), (t_2), (t_3),$ and $(T_{\mathrm{test}})$ are expressed in the normalized units defined by $(\tau=\alpha \kappa t/5)$. The time step $(\Delta t)$, however, is given in the original, unnormalized simulation time.

\subsubsection{$\Delta t$ convergence}

\begin{figure*}[t!]
    \centering
    \includegraphics[width=1\linewidth]{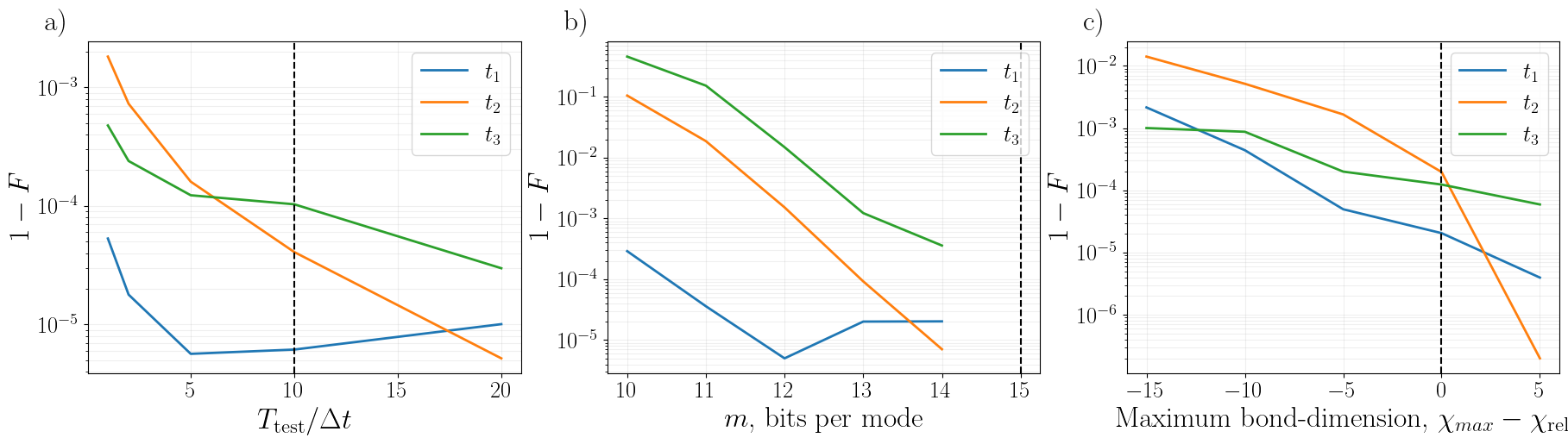}
    \caption{
    Self-convergence analysis of infidelity.
    (a) Infidelity with respect to
     the number of time steps \(N_{\Delta t}\) required to reach
    \(T_{\mathrm{test}}\), where \(N_{\Delta t}=T_{\mathrm{test}}/\Delta t\).
    (b) Infidelity with respect to the number of
    bits per mode \(m\), which controls the quadrature-space resolution at fixed
    spatial support.
    (c)Infidelity with respect to the maximum
    bond dimension \(\chi_{max}\). For convenience, the horizontal axis
    is shifted by the reference bond dimension $\chi_{\mathrm{rel}}$, where \(\chi_{\mathrm{rel}}=30,60,100\) for
    \(t_1,t_2,t_3\), respectively.
    The three curves in each panel correspond to time slices \(t_1=0.34\), \(t_2=0.46\), and \(t_3=0.64\). The infidelity is
    evaluated at the final time of each sub-evolution, \(t_i+T_{\mathrm{test}}\), with
    \(T_{\mathrm{test}}=0.01\).The black dashed lines indicate the parameter values used in the main simulations.}
    \label{Convergence}
\end{figure*}

The self-convergence analysis with respect to the temporal resolution was performed for the following time steps:
\[
\begin{aligned}
\Delta t ={}& 5\times10^{-5},\ 10^{-4}, \\
            & 2\times10^{-4},\ 5\times10^{-4},\ 10^{-3}.
\end{aligned}
\]
The number of bits per mode is 15, and the maximum bond dimension is 30,60 and 100 for $t_1$, $t_2$, and $t_3$, respectively. The reference observable corresponds to $\Delta t_{ref} = 2.5\times 10^{-5}.$

\subsubsection{$\Delta x$ convergence}

The self-convergence analysis with respect to the quadrature-space resolution was performed for the following number of bits per mode (m): 
\[
\begin{aligned}
m=14,13,12,11,10.
\end{aligned}
\] 

The time step is  $\Delta t = 10^{-4}$, and the maximum bond dimension is 30,60 and 100 for $t_1$, $t_2$, and $t_3$, respectively. The reference observable corresponds to the highest-resolution $m_{ref}=15$.

% \begin{figure*}[ht!]
%     \centering
%     \includegraphics[width=1\linewidth]{Figures/Convergence_delta_x.png}
%     \caption
%     {}
%     \label{convergence_delta_t}
% \end{figure*}

%Figure~j+1 demonstrates convergence of the photon-number dynamics with increasing quadrature-space resolution. 
%In particular, the simulation performed with \(m = 14\) differs only negligibly from the reference computation with \(m = 15\), which was used in the main simulations.

\subsubsection{$\chi_{max}$ convergence}

The self-convergence analysis was performed for the following maximum bond dimensions:
\[
\begin{aligned}
\chi_{max} ={}& 15,\ 20,\ 25,\ 30,\ 35
\quad \text{for} \quad t_{\mathrm{1}} = 0.34, \\
\chi_{max} ={}& 45,\ 50,\ 55,\ 60,\ 65
\quad \text{for} \quad t_{\mathrm{2}} = 0.46, \\
\chi_{max} ={}& 85,\ 90,\ 95,\ 100,\ 105
\quad \text{for} \quad t_{\mathrm{3}} = 0.64.
\end{aligned}
\]

The time step and quadrature-space resolution are $\Delta t = 10^{-4}, m=15$, correspondingly.
The reference bond-dimension is $\chi_{\mathrm{ref}}=40, 70, 110$ for $t_1, t_2,t_3$, respectively.

\subsubsection{Results}
Fig.\ref{Convergence} shows self-convergence curves with respect to the main
numerical parameters for the case $\alpha=100$: (a) the number of time steps \(N_{\Delta t}\) required to
reach \(T_{\mathrm{test}}\), corresponding to different values of \(\Delta t\),
with \(N_{\Delta t}=T_{\mathrm{test}}/\Delta t\); (b) quadrature resolution,
controlled by the number of bits per mode \(m\); and (c) maximum bond
dimension $\chi_{max}$. The vertical dashed lines indicate parameter values used in the main simulations.

We observe exponential convergence of the infidelity across both the numerical parameters and the time slices considered. On a logarithmic scale, the curves for $t_2$ and $t_3$ in Fig.\ref{Convergence} are practically straight lines, indicating that the TT algorithm remains stable throughout the evolution.  The parameters adopted for the full simulation ($\alpha=100$, dashed lines in Fig.\ref{Convergence}) are consistent with independent benchmarks, including the maximum pump depletion level, the signal quadrature-squeezing behavior, energy conservation, and a comparative residual analysis against the $\alpha=10$ case. Further details are given in the main text.

\subsection{Implementation details of the MPS time evolution}

The practical implementation of the time evolution for \(\alpha=10\) and
\(\alpha=100\) is summarized in Algorithm~\ref{alg:mps_time_step_alpha10} and
Algorithm~\ref{alg:mps_time_step}, respectively. At each time step, the implicit
equation
\[
U\psi_{t+1}=\psi_t
\]
is solved in the MPS format by one full two-site sweep, consisting of a
left-to-right and a right-to-left pass. The local projected systems are solved
directly. After the two-site tensor \(x\) is obtained, it is factorized back
into two MPS tensors using SVD. For \(\alpha=10\), the SVD decomposition is
truncated according to the fixed relative threshold
\(\epsilon_{\mathrm{SVD}}\), i.e., singular values are discarded based on their
relative contribution to the sum of singular values. For \(\alpha=100\), the
truncation is controlled by the adaptive bond-dimension algorithm.

\begin{algorithm}[t]
\caption{Sweep-based MPS simulation for \texorpdfstring{\(\alpha=10\)}{alpha=10}.}
\label{alg:mps_time_step_alpha10}

\KwIn{
Initial MPS \(\psi_0\); time-step operator \(U=I+i\Delta t H\);
number of time steps \(N_{\Delta t}=900\); SVD truncation threshold
\(\epsilon_{\mathrm{SVD}}=10^{-6}\); initial-guess compression threshold
\(\epsilon_{\mathrm{guess}}=10^{-3}\); maximum bond dimension \(\chi_{\max}=30\).
}

\KwOut{
MPS approximations \(\psi_t\), \(t=1,\ldots,N_{\Delta t}\).
}

\For{\(t=0,\ldots,N_{\Delta t}-1\)}{
    Construct the initial guess for \(\psi_{t+1}\) by compressing
    \(\psi_t\) using SVD truncation: discard singular values \(\sigma_i\)
    satisfying
    \[
        \frac{\sigma_i}{\sum_j \sigma_j}<\epsilon_{\mathrm{guess}} .
    \]

    \For{\(\mathrm{dir}\in\{\mathrm{L}\to\mathrm{R},\,\mathrm{R}\to\mathrm{L}\}\)}{
        \For{each neighboring tensor pair along \(\mathrm{dir}\)}{
            Fix all MPS tensors except the current two-site block\;
            Construct the projected local linear system \(U'x=b\)\;
            Solve the local system directly for the two-site tensor \(x\)\;
            Factorize \(x\) back into two MPS tensors using SVD\;
            Truncate the decomposition by discarding singular values satisfying
            \[
                \frac{\sigma_i}{\sum_j\sigma_j}<\epsilon_{\mathrm{SVD}},
            \]
            while keeping at most \(\chi_{\max}\) singular values\;
        }
    }
}
\end{algorithm}

\begin{algorithm}[t]
\caption{Sweep-based MPS simulation for \texorpdfstring{\(\alpha=100\)}{alpha=100}.}
\label{alg:mps_time_step}

\KwIn{
Initial MPS \(\psi_0\); time-step operator \(U=I+i\Delta t H\);
number of time steps \(N_{\Delta t}=1000\); initial-guess compression threshold
\(\epsilon_{\mathrm{guess}}=10^{-4}\); initial maximum bond dimension
\(\chi_0=\mathrm{bond}(\psi_0)\); upper limit on the maximum bond dimension \(\chi_{\lim}=100\);
bond-dimension increment \(\Delta \chi=10\); residual tolerance
\(\epsilon_{\mathrm{tol}}=0.025\).
}

\KwOut{
MPS approximations \(\psi_t\), \(t=1,\ldots,N_{\Delta t}\).
}

Set the current maximum bond dimension \(\chi_{max} \gets \chi_0\)\;

\For{\(t=0,\ldots,N_{\Delta t}-1\)}{
    Construct the initial guess for \(\psi_{t+1}\) by compressing
    \(\psi_t\) using SVD truncation: discard singular values \(\sigma_i\)
    satisfying
    \[
        \frac{\sigma_i}{\sum_j \sigma_j}<\epsilon_{\mathrm{guess}} .
    \]

    \For{\(\mathrm{dir}\in\{\mathrm{L}\to\mathrm{R},\,\mathrm{R}\to\mathrm{L}\}\)}{
        \For{all neighboring tensor pairs in the current direction}{
            Fix all MPS tensors except the current two-site block\;
            Construct the projected local linear system
            \[
                U'x=b .
            \]
            Solve the local system directly for the two-site tensor \(x\)\;
            Factorize \(x\) back into two MPS tensors using SVD\;
            Truncate the decomposition according to the current maximum bond dimension \(\chi_{max}\)\;
        }
    }

    Compute the normalized residual
    \[
        \epsilon_{\mathrm{res}}^{(t+1)}
        =
        \frac{
        \left\|
        U\psi_{t+1}-\psi_t
        \right\|
        }{
        \left\|
        \psi_t
        \right\|
        } .
    \]

    \If{\(\epsilon_{\mathrm{res}}^{(t+1)}>\epsilon_{\mathrm{tol}}\) and \(\chi_{max}<\chi_{\lim}\)}{
        Increase the bond dimension:
        \[
            \chi_{max} \gets \min(\chi_{max}+\Delta \chi,\chi_{\lim}) .
        \]
        Recompute the same time step with the larger maximum bond dimension\;
    }
}
\end{algorithm}

% \begin{figure*}[ht!]
%     \centering
%     \includegraphics[width=1\linewidth]{Figures/Convergence_bond_limit.png}
%     \caption
%     {}
%     \label{convergence_delta_t}
% \end{figure*}

%Figure~j+2 demonstrates convergence of the photon-number dynamics with increasing bond-dimension truncation limits. The results show that the effect of bond-dimension truncation on the photon dynamics is negligible within the considered parameter range.

\clearpage
%\bibliographystyle{unsrt}
%\bibliography{mybibl}
\end{document}